\documentclass[twocolumn,english,aps,arxiv,superscriptaddress]{revtex4-1}
\usepackage{grffile}
\usepackage[latin9]{inputenc}
\usepackage{color}
\usepackage{babel}
\usepackage{graphicx}
\usepackage{epstopdf}
\usepackage{cancel}
\PassOptionsToPackage{normalem}{ulem}
\usepackage{ulem}
\usepackage[unicode=true, bookmarks=false, breaklinks=false,pdfborder={0 0 1},colorlinks=false] {hyperref}
\usepackage{breakurl}

\begin{document}
%%%%%%%%%%%%%%%%%%%%%%%%%%%%%%%

\title{Structural distortion and incommensurate noncollinear magnetism in EuAg$_4$As$_2$}

\author{Bing Shen$^+$}
\affiliation{Department of Physics and Astronomy and California NanoSystems Institute, University of California, Los Angeles,
CA 90095, USA}

\author{Chaowei Hu$^+$}
\affiliation{Department of Physics and Astronomy and California NanoSystems Institute, University of California, Los Angeles,
CA 90095, USA}
\author{Huibo Cao}
\affiliation{Quantum Condensed Matter Division, Oak Ridge National Laboratory, Oak Ridge, Tennessee 37831, USA}
\author{Xin Gui}
\affiliation{Department of Chemistry, Louisiana State University, Baton Rouge, LA 70803, USA}
\author{Eve Emmanouilidou}
\affiliation {Department of Physics and Astronomy and California NanoSystems Institute, University of California, Los Angeles, CA 90095, USA}
\author{Weiwei Xie}
\affiliation{Department of Chemistry, Louisiana State University, Baton Rouge, LA 70803, USA}
\author{Ni Ni}
\email{Corresponding author: nini@physics.ucla.edu}
\affiliation {Department of Physics and Astronomy and California NanoSystems Institute, University of California, Los Angeles, CA 90095, USA}

\begin{abstract}
Layered pnictide materials have provided a fruitful platform to study various emergent phenomena, including superconductivity, magnetism, charge density waves, etc. Here we report the observation of structural distortion and noncollinear magnetism in layered pnictide EuAg$_4$As$_2$ via transport, magnetization, single crystal X-ray and neutron diffraction data.
EuAg$_4$As$_2$ single crystal shows a structural distortion at 120 K, where two sets of superlattice peaks with the propagation vectors of $q_1=\pm$(0, 0.25, 0.5) and $q_2=\pm$(0.25, 0, 1) emerge. Between 9 K to 15 K, the hexagonal Eu$^{2+}$ sub-lattice enters an unpinned state, with magnetic Bragg reflections pictured as circular-sectors. Below 9 K, it orders in an incommensurate noncollinear antiferromagnetic state with a well-defined propagation wavevector of (0, 0.1, 0.12), where the magnetic structure is helical along the $c$ axis and cycloidal along the $b$ axis with a moment of 6.4 $\mu_B$/Eu$^{2+}$. Furthermore, rich magnetic phases under magnetic fields, large magnetoresistance, and strong coupling between charge carriers and magnetism in EuAg$_4$As$_2$ are revealed.
\end{abstract}
\pacs{}
\date{\today}
\maketitle
%%%%%%%%%%%%%%%%%%%%%%%%%%%%%%%
%%%%%%%%%%%%%%%%%%%%%%%%%%%%%%%
\section{Introduction}

Rare earth elements are characterized with the localized 4$f$ electrons. Often coupled through Ruderman-Kittel-Kasuya-Yosida (RKKY) interaction with conduction electrons, they can develop long-range magnetic order with large local magnetic moments, leading to various magnetic ground states. The intensive studies of the magnetism in rare earth elements and compounds have shaped our understanding of complex magnetism. In rare earth magnet, the magnetism is determined as a result of the interplay of crystal electric field (CEF) anisotropy, the dipole-dipole interaction, the magnetoelastic coupling between the moment and the lattice, as well as the oscillatory RKKY interaction which favors incommensurate periodicity \cite{jensen1991rare}.
Besides these rich interactions which govern the magnetism, other exotic ground states can interplay with the magnetism in rare earth compounds.
Layered Eu$^{2+}$ based pnictides are well-known for being a material platform to study these interplays. For example, EuFe$_2$As$_2$ is an antiferromagnet with a spin density wave phase transition of the FeAs layer at about 190 K and an A-type antiferromagnetism arising from the Eu$^{2+}$ sublattice at around 19 K \cite{ jiang2009metamagnetic,PhysRevB.80.174424}. With K doping on the Eu site, which destroys the Eu magnetism, the SDW transition was suppressed and SC emerged \cite{PhysRevB.78.092406, PhysRevB.80.184514}. Eu magnetism is more persistent with chemical doping of As sites with P\cite{PhysRevB.83.054511,cao2011superconductivity}, and Fe sites with Co\cite{ PhysRevB.83.094520}, or under pressure up to 6 GPa\cite{matsubayashi2011pressure}, leading to coexistence of ferromagnetism and superconductivity.

Recently, we have studied a layered pnictide SrAg$_4$As$_2$, where a structural distortion is observed at around 110 K and quantum oscillation reveals the existence of small Fermi pockets with light effective mass and unexpected high mobility \cite{SrAg4As2}. In this paper, we present our studies of its magnetic analog, EuAg$_4$As$_2$, where the Eu$^{2+}$ magnetic ions have a half-filled 4$f$ shell and form a hexagonal sublattice. At room temperature, it crystallizes in the centrosymmetric trigonal CaCu$_4$P$_2$ structure with the space group of $R\bar 3 m$ \cite{STOYKO2012325}. 
The crystal structure can be taken as inserting an additional itinerant Ag$_2$ layer within the Ag$_2$As$_2$ layer of the trigonal CaAl$_2$Si$_2$-type stacking of -Eu-Ag$_2$As$_2$-Eu- (Fig. 5(a)).
Unlike in the CaCu$_4$P$_2$ where the Cu$_2$ layer locates on one unique fully occupied site, in EuAg$_4$As$_2$, the itinerant Ag$_2$ layer occupies more than one partially occupied sites, leading to metallic Ag-Ag bonding \cite{STOYKO2012325}. The CaAl$_2$Si$_2$-type EuTM$_2$Pn$_2$ (TM = Mn, Zn, Cd and Pn = As and Sb), which are the trigonal version of the 122 pnictide family, were all reported to be antiferromagnetic (AFM) somewhere under 20 K and many are found with excellent thermoelectric properties \cite{1742-6596-391-1-012015,schellenberg2010121sb,PhysRevB.94.014431,dahal2018spin,weber2006low,wang2016anisotropic,toberer2010electronic,zhang2008new}. Antiferromagnetism was also revealed at around 15 K in EuAg$_4$As$_2$ in a study by susceptibility and M\"ossbauer spectroscopy measurement \cite{GERKE201365}, however, very little is known for this compound. Using a combination of transport, single crystal X-ray and neutron diffraction measurements, we revealed the coexistence of a high temperature structural distortion and a low temperature incommensurate noncollinear antiferromagnetism in EuAg$_4$As$_2$.

\section{Single crystal growth and experimental methods}
Single crystals of EuAg$_4$As$_2$ were synthesized using Ag$_2$As as the self-flux. Ag$_2$As precursor was made by the solid state reaction of the stoichiometric ratio of Ag and As powders. The mixture was sealed in a quartz tube, slowly heated to 600 $^{\circ}$C, stayed for 10 hours, and then heated up to and held at 850 $^{\circ}$C for another 10 hours before being cooled to room temperature.
Eu pieces were first scraped and arc-melted to remove oxidization layers. The precursor Ag$_2$As and the Eu pieces were mixed at a molar ratio of 4:1 and placed in an alumina crucible which was then sealed under vacuum in a quartz tube. The tube was heated up to 1100 $^{\circ}$C, stayed for 3 hours, and cooled at a rate of 5$^{\circ}$C/hour to 750$^{\circ}$C. Then the single crystals were separated from the flux using a centrifuge. Several sizable plate-like single crystals with typical dimensions of $5\times4\times2 mm^3$ were obtained. A picture of a single crystal is shown in the inset of Fig. 2(a).

Powder X-ray diffraction data using a PANalytical Empyrean diffractometer (Cu K$\alpha$ radiation) were collected to confirm the phase. Multiple pieces of single crystals ($\sim$20$\times$30$\times$30 $\mu m^3$) were picked up and mounted on the tips of Kapton loop to examine the structure. The data were collected on a Bruker Apex II X-ray diffractometer with Mo radiation K$_{\alpha1}$ radiation ($\lambda$ = 0.71073 \AA) and the temperature ranges from 100 K to 300 K. The SHELXTL package was employed to solve the crystal structure using the direct methods and full-matrix least-squares on F$^2$ models \cite{sheldrick2015crystal}. The incommensurate vectors were examined using JANA 2006 \cite{le1988ab,petvrivcek2014crystallographic}.

The temperature dependence of the resistivity and the specific heat was measured in a Quantum Design (QD) DynaCool Physical Properties Measurement System (DynaCool PPMS) from 300 K to 2 K. The temperature dependence of the magnetization was measured in a QD Magnetic Properties Measurement System (MPMS3).

In order to determine the magnetic structures, single crystal elastic neutron diffraction was performed at the HB-3A four circle diffractometer with the neutron wavelength of 1.550 $\AA$ from a bent perfect Si-220 monochromator at the High Flux Isotope Reactor (HFIR) at the Oak Ridge National Laboratory (ORNL)\cite{Chakoumakos:ko5139}. Representational analysis with the SARAh software was performed to determine the possible magnetic symmetries.\cite{Sarah} The nuclear and magnetic structures were refined with the FullProf Suite software\cite{RODRIGUEZCARVAJAL199355}.

\section{EXPERIMENTAL RESULTS}

\subsection{Superlattice peaks revealed by single crystal X-ray diffraction}

\begin{figure}
  \centering
  \includegraphics[width=3.5in]{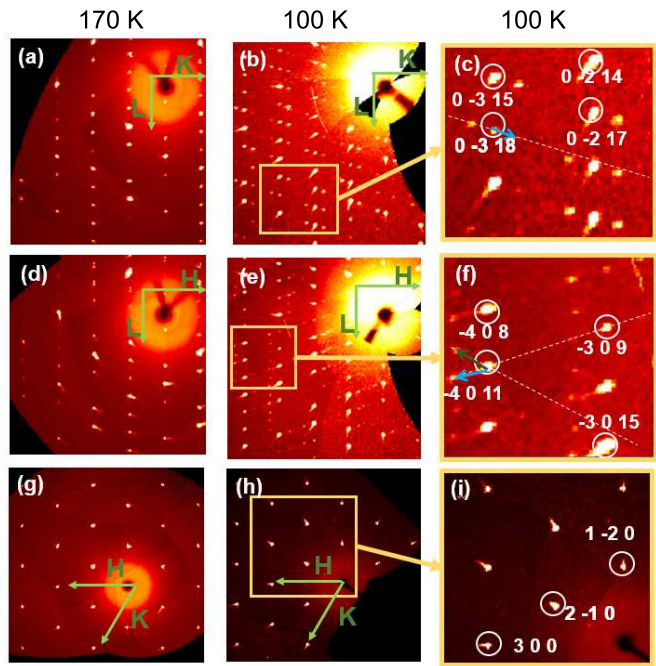}
  \caption{Single crystal X-ray diffraction precession image of EuAg$_4$As$_2$: (a), (d) and (g): the $(0kl)$, $(h0l)$ and $(hk0)$ planes in the reciprocal lattice at 170 K.
 (b), (e) and (h): the $(0kl)$, $(h0l)$ and $(hk0)$ planes in the reciprocal lattice at 100 K. (c), (f) and (i): the zoom-in plot of the selected area of
 (b), (e) and (h).
 The blue and olive arrows indicate the prorogation vectors. Some main Bragg peaks are indexed.
  }
  \label{fig:Fig1}
\end{figure}

The crystal lattice at room temperature is consistent with what the reference reported\cite{STOYKO2012325}. Figure 1 shows the single crystal X-ray precession images of the $(h0l)$, $(0kl)$ and $(hk0)$ planes generated with $2\theta$ = 65$^\circ$ at 170 K and 100 K, respectively. Space group $R\bar 3 m$ requires a reflection condition of $-h+k+l=3n$. At 170 K, in all three planes, only the points satisfying this condition are observed, as shown in Figs. 1(a), (d) and (g). As the temperature decreases to 100 K, additional superlattice peaks arising from the modulation reflections are discernible on the $(0kl)$ and $(h0l)$ planes, which are highlighted in Figs. 1(b) and (e). No detectable superlattice peaks are observed in the $(hk0)$ plane as shown in Fig. 1(h) and the zoom-in plot in Fig. 1(i). Fig. 1(c) presents the close examination of the selected region of the $(0kl)$ plane. The blue arrow connects a main nuclear peak and a superlattice peak next to it, and the associated propagation vector is $q_1=\pm$(0,0.25,0.5). Fig. 1(f) shows the close examination of the selected region of the $(h0l)$ plane. In addition to the symmetry-related vectors of $q_1$ at (-0.25,0,0.5) marked by the blue arrow, the other set of superlattice peaks appears with a commensurate vector $q_2=\pm$(0.25,0,1), which is marked by the olive arrow. The emergence of these superlattice peaks suggests the existence of a structural distortion at low temperature.

The attempt to solve the crystal structure with superlattice model using all X-ray peaks at 100 K yielded a significantly high R value ($\sim$50\%), suggesting the failure of the refinement. Therefore, to get an idea on the main crystal structure at 100 K, only the main X-ray nuclear peaks are used and all superlattice peaks are excluded in the refinement. The refinement suggests that the main crystal structure at 100 K remains the same as the one at 170 K. Tables I and II summarize the crystal structures of EuAg$_4$As$_2$ at 300 K and 100 K based on the main X-ray diffraction peaks and further investigation is needed to solve the low temperature modulated crystal structure.
Unlike the crystal structure determined in Ref. \cite{STOYKO2012325}, in our single crystalline EuAg$_4$As$_2$, the Ag$_2$ layer located on two partially occupied Ag sites, instead of three. This is confirmed based on the refinement of our single crystal neutron diffraction data. The discrepancy may come from the different synthetic processes used in making EuAg$_4$As$_2$.

\begin{table}[t]
\centering
\caption{Single crystal crystallographic data for EuAg$_4$As$_2$ at 100(2) and 300(2) K. To obtain the reliable refinement, the reflection peaks from superstructure have been excluded.}
\begin{tabular}{ccc} \hline
Formula & EuAg$_{3.9(2)}$As$_2$ & EuAg$_{3.9(3)}$As$_2$ \\ \hline
Temperature (K) & 100(2) K & 300(2) K \\
F.W. (g/mol); & 720.34 & 721.41 \\
Space group & ~~~$R\bar 3 m$(No.166)~~~ & ~~~$R\bar 3 m$(No.166)~~~ \\
a (\AA) & 4.514(1) &4.543(2) \\
c (\AA) & 23.554(5) & 23.704(10) \\
V (\AA$^3$) & 415.7(2) & 423.7(4) \\
Extinction Coefficient & 0.00027(5) & 0.0011(1) \\
$\theta$ range (deg) & 2.594- 36.974 & 2.578-36.978 \\
Rint & 2447; 0.0623 & 2375; 0.0781 \\
No. parameters & 21 & 21 \\
R1; wR2 (all I) & 0.0290; 0.0353 & 0.0334; 0.0455 \\
Goodness of fit & 1.078 & 1.051 \\
Diffraction peak & 2.653; -2.046 & 2.661; -2.092 \\\hline
\end{tabular}
\label{tab.1}
\end{table}

\begin{table}[t]
\centering
\caption{Atomic coordinates and equivalent isotropic displacement parameters of EuAg$_4$As$_2$ at 100(2) K and 300(2) K. Wkf is the Wyckoff positions. U$_{eq}$ is defined as one-third of the trace of the orthogonalized U$_{ij}$ tensor (\AA$^2$).}
\begin{tabular}{ccccccc}
\multicolumn{6}{c}{\textbf{100(2) K}}\\\hline
Atom & Wkf. & Occ. & x & y & z &U$_{eq}$ \\ \hline
Eu & 3a & 1 & 0 & 0 & 0 &0.0054(2) \\
Ag1 & 6c & 0.63(3) & 0 & 0 & 0.1466(2) & 0.0158(8) \\
Ag2 & 18h & 0.10(1) & 0.743(5) & 0.257(5) & 0.1797(8) &0.044(7) \\
Ag3 & 6c & 1 & 0 & 0 & 0.4300(1) & 0.0134(2) \\
As & 6c & 1 & 0 & 0 & 0.2588(1) & 0.0069(2)\\ \hline
\multicolumn{6}{c}{\textbf{300(2) K}}\\\hline
Atom & Wkf. & Occ. & x & y & z &U$_{eq}$ \\
Eu & 3a & 1 & 0 & 0 & 0 &0.0133(2) \\
Ag1 & 6c & 0.65(4) & 0 & 0 & 0.1464(4) &0.032(1) \\
Ag2 & 18h & 0.10(1) & 0.74(1) & 0.26(1) & 0.180(2)& 0.072(15)\\
Ag3 & 6c & 1 & 0 & 0 & 0.4297(1) & 0.0257(2) \\
As & 6c & 1 & 0 & 0 & 0.2586(1) & 0.0144(2)\\ \hline
\end{tabular}
\label{tab.2}
\end{table}

\subsection{Physical Properties}
\begin{figure}
  \centering
  \includegraphics[width=3.6in]{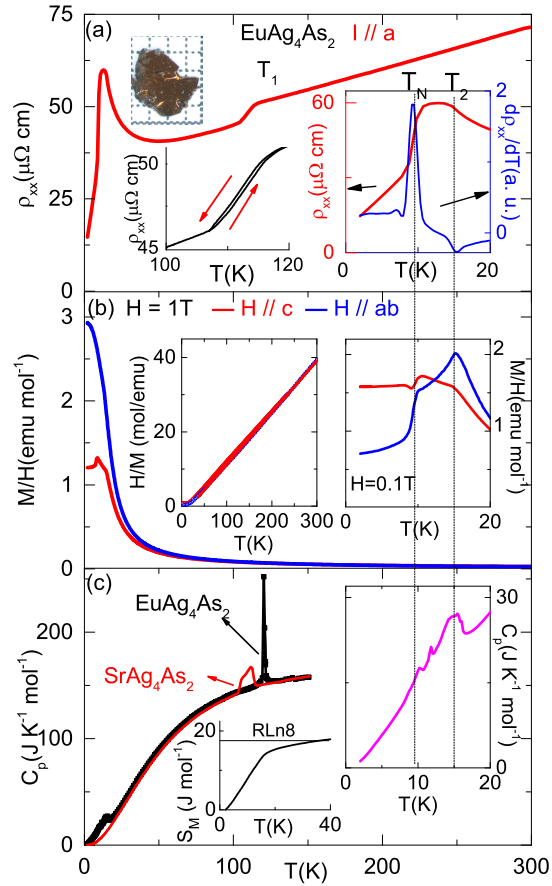}
  \caption{Bulk physical properties of EuAg$_4$As$_2$: (a) The temperature dependent resistivity from 2 K to 300 K. Top left: A single crystal of
  EuAg$_4$As$_2$ against 1-mm scale. Left Inset: The zoom-in resistivity from 100 K to 120 K upon warming and cooling. Right Inset: The zoom-in resistivity and the first derivative of the resistivity from 2 K to 20 K.
  (b) The temperature dependent $M/H$ from 2 K to 300 K with $H//c$ and $H//ab$ and $H=1$ T. Left Inset: Temperature dependent $H/M$ with the Curie-Weiss fit for $H//c$ and $H//ab$. Right Inset: $M/H$ at $H=0.1$ T from 2 K to 20 K with $H//c$ and $H//ab$. (c) The temperature dependent specific heat of EuAg$_4$As$_2$ from 2 K to 150 K. The specific heat of SrAg$_4$As$_2$ is used as a nonmagnetic reference. Left Inset: The entropy release from 2 K to 40 K. Right Inset: The zoom-in specific heat from 2 K to 20 K.
  }
  \label{fig:Fig2}
\end{figure}

Figure 2 summarizes the temperature dependence of the resistivity, susceptibility and specific heat of the EuAg$_4$As$_2$ single crystals. The resistivity, shown in Fig. 2(a), was measured with current applied along the $a$ axis and shows metallic behavior, with a resistivity drop near T$_1 \sim$ 120 K and a sharp peak below 20 K. The left inset shows the zoom-in resistivity around 120 K, which has a slight hysteresis with $\Delta T \approx$1.1 K. Combined with the single crystal X-ray diffraction data in Figure 1, this resistivity drop is associated with the structural distortion and suggests the transition temperature $T_{1} \sim 120$ K. A closer examination of the resistivity data below 20 K is shown in the right inset of Fig. 2(a). Two peaks can be seen in the first derivative of resistivity (blue curve), one around 9 K and the other around 15 K. As will be discussed in more details later, the two peaks are closely associated with the onset of magnetic ordering. The resistivity upturn from 50 K to 15 K suggests strong spin fluctuation in this temperature region, which results in the enhanced spin scattering and thus a resistivity increase. When it enters the magnetic state below 15 K, the resistivity drops sharply due to the loss of spin scattering.

Temperature dependent anisotropic magnetic susceptibility $M/H$ with $H // c$ and $H // ab$ taken at 1 T is shown in Fig. 2(b). Strong magentic anisotropy can be observed. The zoom-in $M/H$ below 20 K taken at 0.1 T is shown in the right inset of Fig. 2(b) to be better compared with transport data under zero field. Two sharp slope changes can be clearly seen at 9 K and 15 K, which agree well with the magnetic phase transition temperatures determined from the first derivative of resistivity shown in the right inset of Fig. 2(a). A Curie-Weiss (CW) fit is made on the $H/M$ data taken at 1 T using $H/M=C/(T-\Theta)$, where $\Theta$ is the Weiss temperature and $C$ is the Curie constant, being related to the effective moment $\mu_{eff}$ by $\mu_{eff} \approx \sqrt {8C}$. The fits of the $H/M$ data with $H//ab$ using the data from 50 K to 300 K and $H//c$ with the data from 50 K to 200 K are presented in the left inset of Fig. 2(b). The $\mu_{eff}$ is 7.73(1)$\mu_B$/Eu and 7.60(1)$\mu_B$/Eu for $H//c$ and $H//ab$ respectively. Both and their poly-crystalline average of 7.65(1) $\mu_B$/Eu are slightly smaller than 7.94$\mu_B$/Eu, the theoretical value of Eu$^{2+}$. This may suggest that a small percentage of nonmagnetic Eu$^{3+}$ atoms is present, consistent with the existence of Ag deficiency from the refinement of the diffraction data. The Weiss temperatures are $\Theta_{ab}$ = 19.4(1) K and $\Theta_{c}$ = 11.0(1) K, and the average $\Theta_{avg}$ = 16.6(1) K is close to the poly-crystalline average of 17.6 K in \cite{GERKE201365}. The positive Weiss temperature suggests dominant ferromagnetic interaction. Considering that the envelope of the low temperature $M/H$ indicates antiferromagnetism, this may suggest a complex magnetic structure here. The absence of feature of the 120 K transition in the $M/H(T)$ again agrees that this transition has a non-magnetic origin. 

A sharp specific heat anomaly is observed at around 120 K in Fig. 2(c), combined with the hysteresis shown in the left inset of Fig. 2(a), indicating the structural distortion is of the first-order type. The zoom-in plot of specific heat below 20 K shows multiple anomalies with the largest entropy release at 15 K, suggesting complex low temperature magnetic transitions \cite {bud2020pressure}. The entropy release is consistent with RLn8, the one for Eu$^{2+}$.

To further investigate the magnetism and charge transport, we performed anisotropic isothermal magnetization and  magneto-transport measurements with $H // c$ and $H // ab$ at 2 K, as shown in Figure 3. In both directions with applied field, multiple transitions are observed.

\begin{figure}
  \centering
  \includegraphics[width=3.4in]{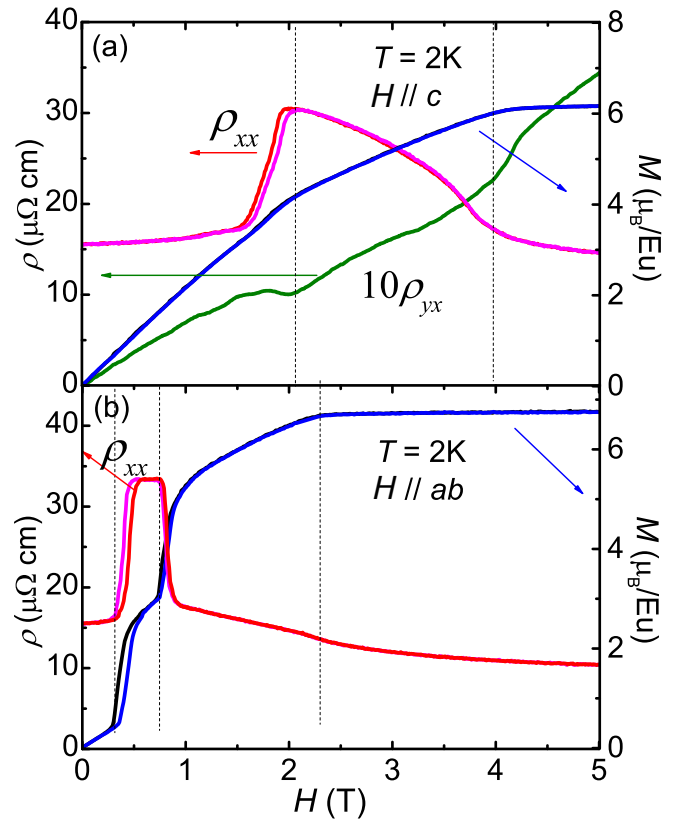}
  \caption{An isotropic isothermal magnetization and Magnetotransport of EuAg$_{4}$As$_2$. (a) Transverse resistivity $\rho_{xx}$, Hall resistivity $\rho_{yx}$, and magnetization measured under with the magnetic field parallel to $c$ at 2 K. (b) Transverse resistivity $\rho_{xx}$ and magnetization measured under the magnetic field parallel to $ab$ at 2 K. The dashed lines in (a) and (b) mark the fields of the transitions. 
  }
  \label{fig:Fig3}
\end{figure}

Fig. 3(a) presents transverse resistivity $\rho_{xx}$, Hall resistivity $\rho_{yx}$, and magnetization $M(H)$ under $H//c$. $M(H)$ shows one slope change before it saturates. At 1.6 T where the first metamagnetic transition occurs, the resistivity goes through a sudden increase with a magnetoreisistance of around 100\%. The presence of hysteresis indicates that the transition is first-order in nature. Above 1.6 T, $\rho_{xx}$ gradually decreases until it reaches the saturation field of around 4 T, which is indicated by the flattened magnetization curve with a saturated moment of 6.2 $\mu_B$/Eu. Beyond saturation, the resistivity reduces very slowly and becomes almost flat. On the other hand, the hall resistivity $\rho_{yx}$ is overall linear except for the intermediate phase around to 1.8 T to 4 T, where the mark of anomalous hall effect is evident. 

When $H//ab$, critical fields for metamagnetic transitions and saturation are lower than those when $H//c$, as shown in Fig. 3(b), suggesting that $ab$ plane is the easy plane. $M(H)$ shows two step transitions before it saturates, indicating more complicated spin reorientations. The first transition occurs at around 0.35 T, where both $M$ and $\rho_{xx}$ have a step-like jump with the magnetoresistance around 100\%. The intermediate phase remains stable between 0.5 T to 0.75 T with a flat $\rho_{xx}$(H) and a mild slope in $M(H)$. EuAg$_{4}$As$_2$ then undergoes a second transition around 0.8 T. At the second transition, $\rho_{xx}$ quickly drops and $M(H)$ has a second sharp increase. Beyond the sharp edge, $M(H)$ slowly increases until the saturation is reached at 2.3 T, which shows up on $\rho_{xx}$ as a mild slope change. Hysteresis is observed for both steps, suggesting first-order phase transitions. 

The multi-step change in $M$, $\rho_{xx}$ and $\rho_{yx}$ suggests the complexity of the magnetic structure and the reorientation of it under fields.

\subsection{Incommensurate noncollinear magnetism}

\begin{figure}
  \centering
  \includegraphics[width=3.4in]{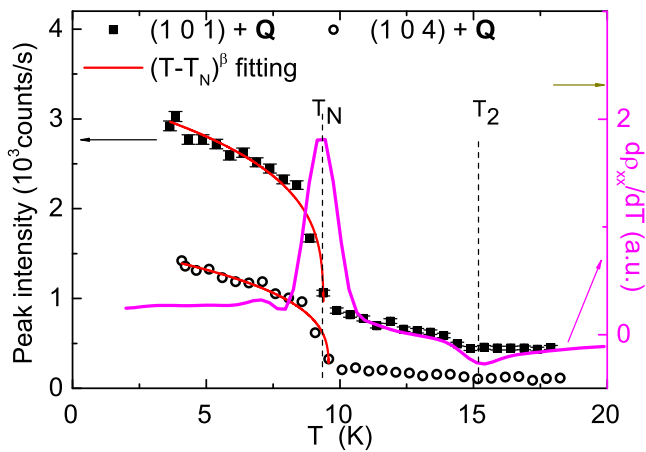}
  \caption{Magnetic order parameter plot of EuAg$_4$As$_2$.  Black symbols: The intensity of the magnetic Bragg peaks (1, 0, 1)+$\textbf{k}_m$ and (1, 0, 4)+$\textbf{k}_m$,
  where $\textbf{k}_m$ is the magnetic prorogation vector (0, 0.1, 0.12).
 Pink Curve: The temperature dependent d$\rho_{xx}$/d$T$. The dashed vertical lines mark the transition temperatures.
  }
  \label{fig:Fig4}
\end{figure}

As we have shown in Figures 2 and 3, rich magnetic phases are present in EuAg$_4$As$_2$, to reveal the nature of the low temperature phase transitions, single crystal neutron diffraction data on EuAg$_4$As$_2$ were collected using HB-3A single crystal neutron diffractometer at HFIR at ORNL under zero magnetic field. Magnetic peaks are observed. The magnetic propagation vector is determined to be incommensurate with $\textbf{k}_m=(0, 0.1, 0.12)$ and used for collecting magnetic reflections and solving the magnetic structure. Figure 4 shows the temperature dependent (1, 0, 1)+$\textbf{k}_m$ and (1, 0, 4)+$\textbf{k}_m$ magnetic peak intensity. As a comparison, temperature dependent d$\rho_{xx}$/d$T$ is also plotted. Upon cooling, both magnetic order parameter (OP) curves show a sudden increase near 9 K, consistent with the d$\rho_{xx}$/d$T$ data shown in pink in Figure 4, which suggests a long-range antiferromagnetic ordering at $T_N$ = 9 K. By fitting the magnetic intensity below $T_N$ with $(T-T_N)^\beta$, we determined the critical exponent to be $\beta=0.34$, close to the expected value of 0.367 from the Heisenberg model. Although slight piece-to-piece variations are present, besides the rapid intensity increase in the OP plot at 9 K, one subtle slope change of the (1, 0, 1)+$\textbf{k}_m$ OP curve is discernible at 15 K, being in good agreement with those seen in the d$\rho_{xx}$/d$T$.

The diffraction lattice of the magnetic phase collected at 4 K is well-defined. Unlike in the single crystal X-ray diffraction data, we did not observe the superlattice peaks in the single crystal neutron diffraction. This may be due to the fact that the superlattice peaks are too weak for the neutron diffraction to detect. In the magnetic structure refinement carried out using the $R\bar1$ symmetry, only the Eu$^{2+}$ magnetic ions were considered. The refined magnetic structure is shown in Figure 5. It is a complex noncollinear magnetic structure. At 4 K, the spin of Eu$^{2+}$ lies on the $ab$ plane and the moment of each Eu$^{2+}$ site was found to be 6.4$\mu_B$, as shown in Fig. 5(a). The small values of the propagation vector $\textbf{k}_m=(0, 0.1, 0.12)$ along both $a$ and $c$ directions indicate long periodicity along these two directions. Fig. 5(b) shows the demonstration of the magnetic structure of  Eu$^{2+}$ ions in the $ac$ plane. The Eu$^{2+}$ spin rotates around the $c$ axis, suggesting a helical arrangement. Fig. 5(c) shows a cartoon of the magnetic structure in the $ab$ plane. The Eu spin rotates along the $b$ axis, suggesting a cycloidal arrangement. Therefore, the periodicity resembles that of a typical helical structure with a period of about $8c$ along the $c$ axis, which is nearly 20 nm, and a cycloidal structure with a period of $10b$ along the $b$ axis, which is nearly 4.5 nm.

\begin{figure}
  \centering
  \includegraphics[width=3.4in]{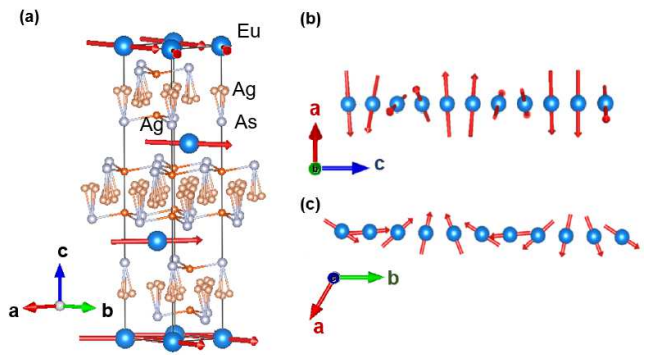}
  \caption{The noncollinear magnetic structure of EuAg$_4$As$_2$. (a) the structure in one unit cell at 4 K. Blue sphere: Eu. Grey sphere: As. Orange sphere: Fully occupied Ag. Light orange sphere: Partially occupied Ag. (b)-(c) Periodicity beyond one unit cell along the $c$ and and $b$ direction, respectively. Only Eu$^{2+}$ ions are shown.  }
  \label{fig:Fig5}
\end{figure}

The evolution of magnetism can be visualized in details by examining the diffraction image in the reciprocal space. The overall intensity of the scan is integrated within each solid angle and projected on a sphere centered at (2, 0, 1). Figs. 6(a) and (b) show the image of the diffraction sphere sliced in the reciprocal $ab$ plane with $\Delta k_z=0.0026$ \AA $^{-1}$ and -0.0026 \AA$^{-1}$, respectively. The coordinates are relative to (2, 0, 1). At 4 K, as shown in Figs. 6(a) and (b), in total six ellipsoid-like magnetic Bragg peaks are observed around (2, 0, 1), indicating the existence of long-range antiferromagnetic ordering. The six peaks are located $|\textbf{k}_m|$ away from the nuclear peak and can be related by symmetry operations in the point group of $D_{3d}$. Fig. 6(c) shows the mapping of the diffraction sphere with the latitude (azimuthal angle) $\theta$ and longitude (polar angle) $\phi$. The three peaks in Figs. 6(a) and (b) correspond to the bottom and top three peaks in Fig. 6(e), respectively. The slight asymmetries of the peak shapes and intensities may be due to the different extent of absorption along different directions.

\begin{figure}
  \centering
  \includegraphics[width=3.4in]{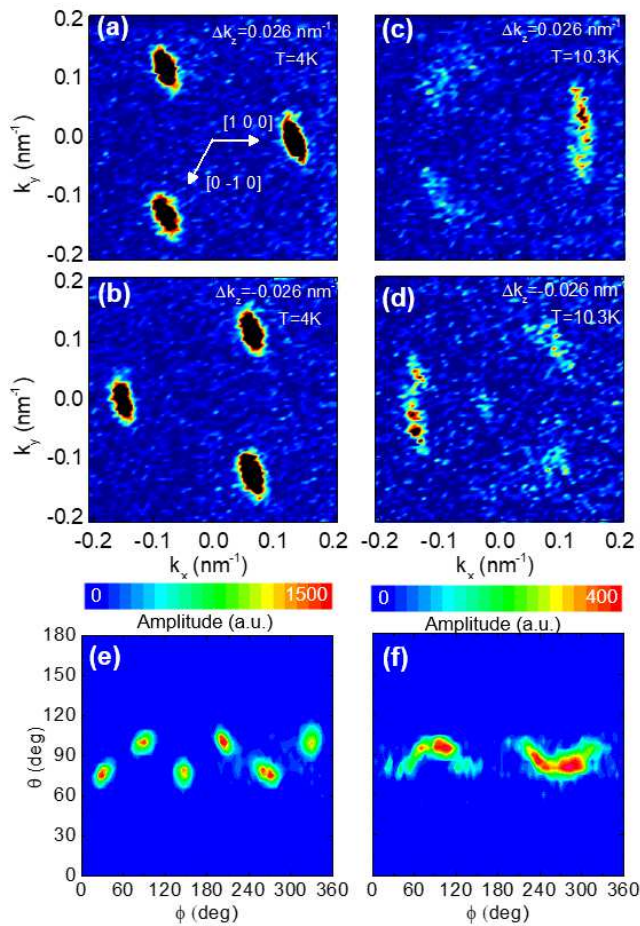}
  \caption{
  Single crystal neutron diffraction of EuAg$_4$As$_2$ near the (2, 0, 1) nuclear peak:
(a)-(b) Slices at $\Delta k_z$=$\pm$0.026 nm$^{-1}$ taken at 4 K. Coordinates are given relative to the (2, 0, 1) nuclear peak. (c)-(d) Slices at $\Delta k_z$=$\pm$0.026 nm$^{-1}$ taken at 10.3 K. (e)-(f) Cylindrical mapping of the magnetic reflection, with integrated magnetic diffraction intensity along each azimuthal angle $\theta$ and polar angle $\phi$ at 4 K and 10.3 K, respectively.
  }
  \label{fig:Fig6}
\end{figure}
Figs. 6(c), (d) and (f) contain the same slices and angular projections of the diffraction sphere at 10.3 K as Figs. 6(a), (b) and (e) do. Diffraction intensities in Figs. 6(c) and (d) can be observed at approximately the same locations as the six peaks in Figs. 6(a) and (c), but they become circular-sector-like magnetic Bragg bands with the same magnitude of the propagation vector. As shown in Fig. 6(f), upon warming, six distinct ellipsoids in Fig. 6(e) evolve into two U-shaped bands, and the symmetry of the magnetic Bragg peaks is lowered to $C_i$. 

\section{DISCUSSION}
In EuAg$_4$As$_2$, since Eu$^{2+}$ ions form hexagonal lattices, frustration may play a role in the incommensurate periodicity. However, the frustration parameter measured by $|\Theta|/T_N$ is around 1.1 so the magnetic frustration may not be the dominant factor. Since Eu$^{2+}$ has half-filled 4$f$ orbitals, the lack of orbital moment leads to very weak CEF anisotropy, which is indeed suggested by isotropic Curie-Weiss curves shown in the left inset of Fig. 2(b). Furthermore, the dipole-dipole interaction is usually one or two orders smaller than the RKKY interaction. Therefore, the RKKY interaction dominates here, which readily explains the existence of the incommensurate magnetic periodicity. Incommensurate noncollinear magnetism has been found in Eu-containing materials. For example, the tetragonal EuCo$_2$As$_2$ and EuCo$_2$P$_2$ adopt incommensurate magnetic structure in the $ab$ plane with a helical axis along the $c$ direction \cite{PhysRevB.95.184404,tan2016transition,reehuis1992neutron}. Similar AFM structure was also observed in GdSi, and GdCo$_2$Ge$_2$ and GdNi$_2$Ge$_2$ \cite{feng2013incommensurate,good2005magnetic,welter2001structural,islam1999effects} where Gd$^{3+}$ ions share a similar electronic state as Eu$^{2+}$.

Although it is clear that long range incommensurate non-collinear AFM order develops below 9 K, the nature of the magnetic order between 9 K and 15 K needs some clarification. As shown in Fig. 6(c) and (d), the well-defined six ellipsoid-like magnetic Bragg peaks at 4 K become circular-sector-like magnetic bands at 10.3 K while maintaining the distance from the nuclear peak. This may suggest a unpinned (also called partial or disordered) non-collinear magnetic state \cite{karube2018disordered, sarkis2019partial}. It is interesting to see that majority of the entropy releases at around 15 K when EuAg$_4$As$_2$ enters this disordered magnetic state which shows long-range ordering feature in the recent M\"{o}ssbauer measurement \cite{ryan2019magnetic}. Therefore, at 15 K, EuAg$_4$As$_2$ undergoes an unpinned noncollinear magnetic state, and when the thermal excitation is lower at 9 K, the order becomes fully pinned, giving the ordinary ellipsoid magnetic reflection. Questions remain to be answered about the driving force of the unpinning. Frustration or chemical disorder may contribute to the unpinned state. As we just discussed that the frustration in EuAg$_4$As$_2$ is not strong, this may suggest that chemical disorder plays a significant role here. Indeed, EuAg$_4$As$_2$ is sensitive to the synthesis condition \cite{bud2020pressure}. Further investigation on doped EuAg$_4$As$_2$ may help clarify this.

\section{Conclusion}
In conclusion, EuAg$_4$As$_2$ undergoes a first-order structural distortion at around 120 K, where superlattice peaks are observed with two independent propagation vectors of (0.25, 0, 0.5) and (0, 0.25, 1). It enters an unpinned incommensurate antiferromagnetic state at 15 K. Below 9 K, the magnetism develops into a long-range incommensurate state with a definite propagation vector of (0, 0.1, 0.12) where Eu$^{2+}$ spins lie in ab plane and order helically along the $c$ axis and cycloidally along the $b$ axis with a moment of 6.4 $\mu_B$/Eu$^{2+}$. EuAg$_4$As$_2$ shows several metamagnetic transitions under magnetic fields, leading to multi-step magnetotransport with the largest magnetoresistance of around 100\% at 2 K, revealing rich magnetic phases and strong coupling between magnetism and charge carriers.

\section*{Acknowledgments}
Bing Shen and Chaowei Hu contributed equally. We thank Prof. Paul. C. Canfield, Dr. Sergey bud$^\prime$ko and Prof. D. H. Ryan for useful discussion and encouragement to finish and publish this work. Work at UCLA was supported by the U.S. Department of Energy (DOE), Office of Science, Office of Basic Energy Sciences under Award Number DE-SC0011978. Work at ORNL's High Flux Isotope Reactor was sponsored by the Scientific User Facilities Division, Office of Basic Energy Sciences, DOE. Work at LSU was supported by Beckman Young Investigator (BYI) Program.

\section*{Comments}
On June 2, 2020, the paper was accepted by Physical Review Materials.

\medskip
\bibliographystyle{apsrev4-1}
\bibliography{Eu142}
\end{document}